\documentclass[aps,twocolumn]{revtex4}

\usepackage{graphicx}
\usepackage{dcolumn}
\usepackage{bm}
\usepackage{amsmath}
\begin{document}

\newcommand{\bk}{{\bf k}}
\newcommand{\mub}{{\mu_{\rm B}}}
\newcommand{\sD}{{\scriptscriptstyle D}}
\newcommand{\sF}{{\scriptscriptstyle F}}
\newcommand{\sCF}{{\scriptscriptstyle \mathrm{CF}}}
\newcommand{\sH}{{\scriptscriptstyle H}}
\newcommand{\sAL}{{\scriptscriptstyle \mathrm{AL}}}
\newcommand{\sMT}{{\scriptscriptstyle \mathrm{MT}}}
\newcommand{\sT}{{\scriptscriptstyle T}}
\newcommand{\up}{{\mid \uparrow \rangle}}
\newcommand{\down}{{\mid \downarrow \rangle}}
\newcommand{\upt}{{ \langle \uparrow \mid}}
\newcommand{\downt}{{\langle \downarrow \mid}}
\newcommand{\bbar}{{\mid \uparrow, 7/2 \rangle}}
\newcommand{\abar}{{\mid \downarrow, 7/2 \rangle}}
\renewcommand{\a}{{\mid \uparrow, -7/2 \rangle}}
\renewcommand{\b}{{\mid \downarrow, -7/2 \rangle}}
\newcommand{\plus}{{\mid + \rangle}}
\newcommand{\minus}{{\mid - \rangle}}
\newcommand{\ex}{{\mid \Gamma_2^l \rangle}}
\newcommand{\LH}{{{\rm LiHoF_4}}}
\newcommand{\LHx}{{{\rm LiHo_xY_{1-x}F_4}}}

\title{Significance of the hyperfine interactions in the phase diagram
of ${\rm LiHo_xY_{1-x}F_4}$}

\author{M. Schechter and P. C. E. Stamp}

\affiliation{Department of Physics \& Astronomy and Pacific Institute for 
Theoretical Physics, University of British Columbia, Vancouver, British 
Columbia, Canada V6T 1Z1}


\begin{abstract}

We consider the quantum magnet $\rm LiHo_xY_{1-x}F_4$ at $x = 0.167$.
Experimentally the spin glass to paramagnet transition in this system
was studied as a function of the transverse magnetic field and
temperature, showing peculiar features: for example (i) the spin glass
order is destroyed much faster by thermal fluctuations than by the
transverse field; and (ii) the cusp in the nonlinear susceptibility
signaling the glass state {\it decreases} in size at lower
temperature. Here we show that the hyperfine interactions of the Ho
atom must dominate in this system, and that along with the transverse
inter-Ho dipolar interactions they dictate the structure of the phase
diagram. The experimental observations are shown to be natural
consequences of this.

\end{abstract}

\maketitle

$\LHx$ is widely considered to be a model quantum Ising magnet.
The strong easy axis crystal field gives the large spin Ho ion an
Ising character, behaving as a 2-level system when temperature $T
\ll 10$ K; quantum fluctuations between the 2 Ising states are
induced by a tunable transverse magnetic field $H_{\perp}$. If one
then neglects both hyperfine (hf) and transverse dipolar (TD)
interactions, the system is described by a transverse field Ising
model (TFIM).  Experimentally, for $x=1$ the system orders ferromagnetically
at low $T$ and $H_{\perp}$\cite{BRA96}; when $x < 1$ the longitudinal
dipolar interactions become random and frustrated, and for $x=0.167$
it was observed that the system orders in a spin glass (SG) 
state\cite{WER+91,WBRA93}. Much of the recent theoretical interest in
quantum spin glasses\cite{MH93,You97,Sac99} has
been driven by these experiments\cite{WER+91,WBRA93}, and both general
theories on quantum spin glasses (see e.g. Ref. [\onlinecite{CGS00}])
and specific models of the quantum Ising SG\cite{BSC94,CBS94}
are commonly checked against their results.

However, for $x = 0.167$ the hf interaction is larger than
the mean dipolar interaction, and in part (i) of the Letter we show that 
it actually dominates the low-energy
physics in this system. 
The longitudinal hf term forces the two ground doublet states to have a
definite, and opposite, nuclear spin\cite{GWT+01,BGW+04}. Low transverse
magnetic fields ($\mub H_{\perp} \ll \Omega_o$, where $\Omega_o$ is
the gap to the first excited electronic state) cannot then couple or
induce quantum fluctuations between these Ising-like states, but
only renormalize their effective spins (and hence the effective
dipolar interaction). Transverse hf interactions can induce
fluctuations between these states; however, when $\mub H_{\perp} \ll
\Omega_o$ we see that these fluctuations are very weak.
Therefore, at low transverse fields, the low-energy effective
Hamiltonian is given by the classical Ising model, with renormalized
parameters. We thus explain why $g \mu_B H_c \gg kT_c$ in the
experiment \cite{WBRA93}, i.e., why the transverse field coupling
required to destroy the glass is so much larger than the thermal
energy required to do the same: While $T_c$ at zero field is given by
the mean dipolar interaction, the temperature dependence of the
transition line on $H_{\perp}$ is dictated by the larger hf
interaction that controls the renormalization of the effective spin,
and the critical transverse field is controlled by the even larger
energy scale of $\Omega_o$.

To completely understand how fluctuations are switched on by
transverse fields, we show that TD terms must be
included.  By doing this in a simple way we obtain a satisfactory
quantitative agreement with the experimental phase diagram.  We then
consider the diminishing of the cusp in the nonlinear susceptibility
$\chi_3$, found in the experiment\cite{WBRA93} as temperature is
reduced. A natural consequence of our theory is that this may be a
result of the renormalization of the effective spin, rather than a
first order phase transition.

The Hamiltonian describing the $\LHx$ system is given by a sum of
crystal field\cite{GWT+01,CHK+04}, Zeeman, hf, and inter-Ho
interaction terms:
\begin{equation}
H = H_{\rm cf} + H_{\rm Z} + H_{\rm hf} + H_{\rm int} \, . \label{generalH}
\end{equation}

The Ho atom has a total angular momentum of $J=8$, and nuclear spin
$I=7/2$. The Zeeman term $H_Z=- \sum_i g_J \mub \vec{H} \cdot
\vec{J_i}$, and $H_{\rm hf} = A_J \sum_i \vec{I_i} \cdot \vec{J_i} \,$ 
is the hf interaction. $H_{int} = - \sum_{ij}
U_{ij}^{\alpha \beta} J_i^\alpha J_j^\beta$ is dominated by the
dipolar interaction\cite{CHK+04}. The interaction of the external
magnetic field with the nuclear spins is small, and is neglected here.

The crystal field $H_{cf}$ leaves only three electronic states at low energy; 
a low-energy doublet, denoted $\up$ and $\down$ here, and an
excited state $\vert \Gamma_2^l\rangle$ roughly $\Omega_o=10.5$ K above
these. Because of the strong crystal anisotropy the TD and
hf terms are usually neglected. Neglecting the longitudinal
hf interactions as well, and in the presence of transverse
magnetic field $H_{\perp}$, the low-energy effective Hamiltonian for
the system would then be the simple TFIM\cite{BRA96}:
\begin{equation}
  H =  - \sum_{i,j} V_{ij}^{zz} \tau_i^z  \tau_j^z - \Delta_o
(H_{\perp}) \sum_i \tau_i^x \, ,
 \label{IsingH}
\end{equation} 

where $\vec{\tau}_j$ is a Pauli vector describing the two-level
effective electronic spin at spatial position ${\bf r} = {\bf r}_j$,
the $j$-th site, $V_{ij}^{zz} \approx 150 U_{ij}^{zz}$ (with $\vert
V_{ij}^{zz} \vert \sim 0.3$ K for nearest-neighbour Ho spins), and
$\Delta_o$ is the transverse field-induced coupling between the
eigenstates $\up$ and $\down$ of ${\hat{\tau}}^z$. For small fields
$\Delta_o \propto H_{\perp}^2$ coming from the second-order coupling, 
mainly via the state $\vert \Gamma^l_2 \rangle$ [so that $\Delta_o
\sim 9 (\mu_BH_{\perp})^2/\Omega_o$].  In this simple
model quantum fluctuations are already important at fairly small
transverse fields, and we expect a $T=0$ quantum phase transition
(QPT) when $\Delta_o \sim V_o$, where $V_o \equiv \langle
\sum_j{V_{ij}^{zz}} \rangle$ is the average dipolar interaction.

For the undiluted system, $x=1$, it was indeed
shown\cite{BRA96,BD01,CHK+04} that the ferromagnetic paramagnetic (PM) 
transition line is well described by this model, with a low
temperature correction due to hf interactions. However, in the
dilute case $V_o$ is reduced by a factor $x$, so the hf
interaction is much larger than $V_o$.  The model (\ref{IsingH}) is
then inadequate, and we must include the hf interactions from
the beginning. We now introduce the interactions neglected in the
model (\ref{IsingH}) consecutively, emphasizing the significance of each
on the structure of the phase diagram.

\vspace{3mm}

\begin{figure}
\includegraphics[width = \columnwidth]{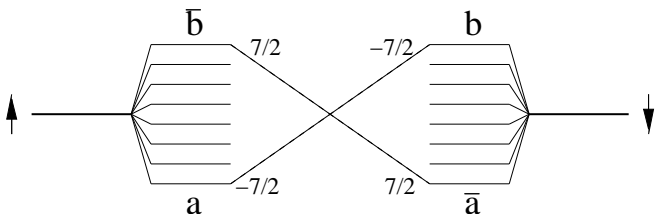}
 \caption{Splitting of the electronic low-energy doublet ($\uparrow$ and
$\downarrow$) due to the longitudinal hyperfine interaction. The
ground state doublet, $a$ and $\bar{a}$ have a definite and opposite
nuclear spin, $\pm 7/2$. Transverse magnetic field couples states
with the same nuclear spin, as is shown by the dashed lines. }
    \label{fignuclearsplitting}
\end{figure}

(i) {\it Hyperfine Interactions}: We first consider the regime $\mub
H_{\perp} \ll \Omega_o$ in which the transverse hf interactions
are negligible.  The longitudinal hf interaction $H_{\rm hf}^{z} =
A_J I^z J^z$ splits each of the states $\up$, $\down$ into an eightfold
multiplet of nearly equidistant levels, with separation $\sim 205 
$mK\cite{GWT+01} between adjacent levels
(Fig.\ref{fignuclearsplitting}). The two lowest energy Ising 
states in zero field are now $a
\equiv \a$ and $\bar{a} \equiv \abar$, and when $H_{\perp} \neq 0$,
these become 
\begin{equation}
\mid + \rangle = c_1 \mid a \rangle + c_2 \mid b \rangle
\hspace{0.5cm} ; \hspace{0.5cm} \mid - \rangle = c_1 \mid \bar{a}
\rangle + c_2 \mid \bar{b} \rangle \; \; ,
\end{equation}
with coefficients $c_1 = \alpha \Delta_o$, and $c_2 = \alpha
[A-\sqrt{A^2+\Delta_o^2}]$, where $\alpha = [\Delta_o^2
+(A-\sqrt{A^2+\Delta_o^2})^2]^{-1/2}$ and $2A \approx 1.4$ K is the
energy difference between the states $a (\bar{a})$ and $b \equiv \b
(\bar{b} \equiv \bbar)$.

The crucial point here is that the transverse magnetic field
$H_{\perp}$ does not couple the relevant Ising doublet states: $a$ and
$\bar{a}$ (it actually couples $a$ to $b$ and $\bar{a}$ to
$\bar{b}$). This immediately invalidates the TFIM (\ref{IsingH}), 
since $H_{\perp}$ no longer induces
fluctuations. The only effect of the transverse field is to
re-normalize the spin: For the state $\mid + \rangle$ one finds
\begin{equation}
\langle \sigma_{+}^z \rangle = \eta \langle
\sigma_{+}^z(0) \rangle  \hspace{1cm} ; \hspace{1cm}
\eta \equiv (c_1^2 - c_2^2) \, ,
 \label{eta}
\end{equation}
and $\langle \sigma_{-}^z \rangle = - \langle \sigma_{+}^z \rangle$.
One can then absorb the renormalization into the dipolar
interaction, and therefore our system reduces at low transverse
fields to the classical Ising model:
\begin{equation}
  H_{\rm eff}^{\parallel} =
- \sum_{i,j} \tilde{V}_{ij}^{zz} s_i^z  s_j^z  \, ,
 \label{IsingClassH}
\end{equation}
where $s$ is a spin half matrix in the space of the states $\mid +
\rangle$ and $\mid - \rangle$ and $\tilde{V}_{ij}^{zz} = \eta^2
V_{ij}^{zz}$.

As a result, the SG paramagnet transition line is given (in
this model, neglecting the later to be shown significant TD 
interactions) by the relation $T_c = \eta^2 T_c(0)$, and the
dependence on the transverse field is through
$\eta(\Delta_o)$. Importantly, this result is independent of the
approximation used to treat the longitudinal dipolar interaction, and
relies only on the assertion that $T_c$ scales linearly with the
interaction.  For $\Delta_o \ll A$ one has from (\ref{eta}) that
$\eta=1-\frac{\Delta_o^2}{2A^2}$, and defining $\epsilon \equiv (T_c -
T)/T_c$ one finds that for $\Delta_o/A, \epsilon \ll 1 $ (i.e., small
$H_{\perp}$ and $T \sim T_c$) the phase transition line $\Delta_c(T)$
obeys the relation $\Delta_c = A \sqrt{\epsilon}$.
For the initial TFIM (\ref{IsingH}) one would
instead get $\Delta_c \sim V_o \sqrt{\epsilon}$. The difference arises
because the hf energy scale $A$ dictates the reduction of the
effective spin (and equivalently the effective dipolar interaction).
At $T=T_c/2$ one finds that $\Delta_c \approx A$, and for $\Delta_o/A
\gg 1$ we have $\eta=A/\Delta_o$, so that $\Delta_c=A \sqrt{(V_o/T)}$.
{\it Thus, as long as the transverse hyperfine interactions are
negligible ($\mub H_{\perp} \ll \Omega_o$) there is no QPT.}

We now turn to the discussion of quantum fluctuations in this system,
and we thus consider the transverse hf coupling $H_{\rm hf}^{\perp}
= A_J (I_+ J_- + I_- J_+)/2$. This term changes the $z$ component of
the nuclear spin by coupling the electronic states $\up$ and $\down$
with (mainly) the state $\ex$ at energy $\Omega_o$, thus inducing
fluctuations between $\plus$ and $\minus$. However, unless $\xi
\equiv \upt \mu_{\rm B} H_{\perp} \mid \Gamma_2^l \rangle \sim
\Omega_o$ (corresponding to a field $H_{\perp} \sim 2$T), these
fluctuations are much smaller than $V_o$, and cannot induce a QPT. 
To see this more quantitatively we diagonalize the
single Ho Hamiltonian in a transverse magnetic field, i.e., we
diagonalize $H = H_{cf} + H_Z + H_{\rm hf}$, 
in the $136$ eigenfunction space (17 crystal field states * 8
nuclear states) using the parameters used in
Ref.~[\onlinecite{CHK+04}]. We then plot in Fig. 2 the results for the
splitting $\tilde{\Delta}(H_{\perp})$ between the two lowest levels
induced by the combination of $H_{\perp}$ and $H_{\rm hf}^{\perp}$.  For
$H_{\perp} \lesssim 3$T the two lowest levels are well separated from
the higher energy levels, and we can replace the classical Ising
Hamiltonian (\ref{IsingClassH}) by an effective Hamiltonian
\begin{equation}
  H =
- \sum_{i,j} \tilde{V}_{ij}^{zz}(H_{\perp},A) s_i^z  s_j^z -
\tilde{\Delta}(H_{\perp},A,\Omega_o) \sum_i  s_i^x \, .
 \label{IsingHeff}
\end{equation} 
We then expect a zero temperature QPT when
$\tilde{\Delta}(H_{\perp}) \rightarrow \Delta_c(H_{\perp})$, such that
$\Delta_c(H_{\perp})\approx V_o(H_{\perp})$. From Fig. 2 we confirm
that this happens when $H_{\perp} \sim 2$T. Note there are now three 
significant energy scales in the problem. $T_c$ is dictated by $V_o$,
while $\Delta_c$ is related to both $A$, which dictates the
renormalization of the effective dipolar interaction, and to
$\Omega_o$.  Interestingly, for any practical dilution $x$ the lower
bound for the critical transverse magnetic field is set by $\Omega_o$,
since only at $\xi \approx \Omega_o$ do quantum fluctuations between the
relevant Ising states become appreciable.

We may summarize our analysis so far by saying that the longitudinal
hf interactions invalidate the usual quantum Ising model for
the $\LHx$ system, instead creating a classical Ising model whose
energetics is determined by the longitudinal dipolar interactions,
renormalized by longitudinal hf interactions. Adding in the
transverse hf interactions brings back an effective transverse
field term $\tilde{\Delta}(H_{\perp})$ into the effective Hamiltonian,
but $\tilde{\Delta}(H_{\perp})$ does not become important until much
higher transverse fields than in the original model (and it switches
on with field in a quite different way from the usual form $\Delta_o
\propto H_{\perp}^2$). We emphasize that
$\tilde{\Delta}$ should be used instead of $\Delta_o$ in analysis of
the phase transition and, in particular, for the determination of the
critical exponents in the system.

Can our simple model approximation explain the experimental phase
diagram? Since according to our analysis up to now, the position of
the phase transition line is mainly dictated by single atom properties
and the hf interaction, we now derive a phase diagram treating
the single Ho exactly, and the inter-Ho interactions by using a 
mean-field (MF) approximation in which each spin feels an interaction 
strength proportional to its average $J_z$, i.e., using the MF 
Hamiltonian
\begin{eqnarray}
H_{\rm MF} \;\;=\;\; H_{\rm cf} &-& \sum_j g_J \mu_B
H_{\perp}J_j^x
\nonumber \\
&+& A_J \sum_j \vec{I_j} \cdot \vec{J_j} + \sum_{j} V_o
\langle J_j^z \rangle J_j^z \, .
 \label{longdipH}
\end{eqnarray}
In Fig. \ref{figdiagram} we plot by dashed line the SG-PM 
transition obtained within this approximation. $V_o$ is
fixed by the experimental value of $T_c$ at zero transverse field.  In
order to compare our results to Fig. 1 of Ref.~[\onlinecite{WBRA93}] 
we plot the phase diagram as function of $T$ and
$\Delta_o$, which is given by the splitting of the electronic levels when
the hf interactions are neglected.

\begin{figure}
\includegraphics[width = \columnwidth]{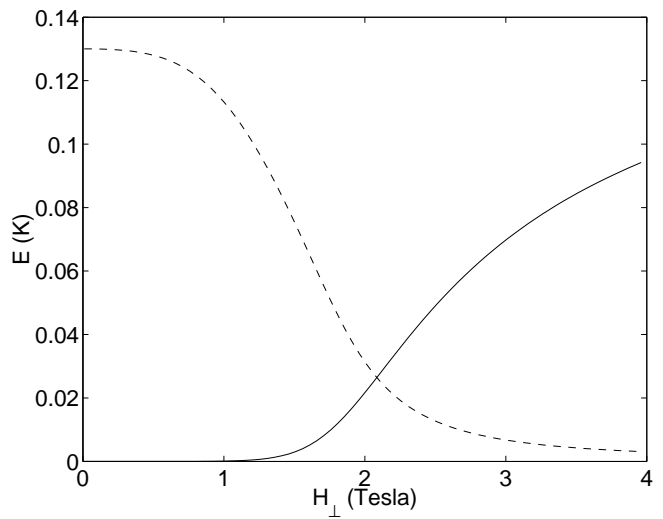}
 \caption{The splitting $\tilde{\Delta}(H_{\perp},A,\Omega_o)$
between the ground and first excited state of the single-ion
Hamiltonian as a function of $H_{\perp}$ (solid
line) calculated by exact diagonalization. This measures the
quantum fluctuations between the states $\plus$ and $\minus$. The
dashed line is the averaged longitudinal dipolar interaction, taking
$V_o=0.13$ for $H_{\perp}=0$ to match the experimental $T_c$ at zero
field. }
    \label{fighfsplitting}
\end{figure}

This mean field result does capture two main features of the
experiments, viz. (i) the much larger energy associated with quantum
disordering of the SG state than with temperature disordering,
and (ii) the roughly straight transition line in the $\Delta_o, T$
plane in most of the parameter regime. However two noticeable
differences are apparent, viz., (i) the experimental $T=0$ quantum
critical point occurs at lower transverse field and (ii) the shape of
the experimental transition line near the zero field transition point
is quite different from the predicted square root behavior. These
differences are {\it not} a result of our MF approximation.
As discussed above, the square root behavior at low transverse fields
is a direct consequence of the renormalization of the spin by the
transverse magnetic field, and is a single atom property.
Furthermore, the experimental graph suggests that the quantum
fluctuations induced by the transverse hf interactions are
significant already at fields of the order of $1$ Tesla, in contrast
with Fig.\ref{fighfsplitting}.

\vspace{3mm}

(ii) {\it Transverse Dipolar Interactions}: The differences just noted
suggest that the TD interactions [e.g., $(\alpha \beta)=(zx)$], 
neglected in the
Hamiltonian (\ref{longdipH}), are also significant in the diluted
$\LHx$ system.  Unlike for $x=1$, where the transverse terms cancel by
symmetry, in the diluted system this is not the case- the TD 
terms add an effective magnetic field at each Ho
location. This field adds both an effective random longitudinal term
at each site\cite{Gin05}, 
and a transverse term which can induce fluctuations even
in the absence of $H_{\perp}$. We consider the latter effect here, and
the former in connection with the nature of the phase transition
below. For $\mu_{\rm B} H_{\perp} \approx V_o$ or larger it would seem
that the induced transverse fields are random and could enhance or
decrease the effective transverse field at each location. However, the
external magnetic field breaks time reversal symmetry, and the
configurations in which the TD interactions enhance
the effective magnetic field at the Ho sites are energetically
favorable, as can easily be seen by comparing the two Ho states $\mid
\uparrow,\rightarrow \rangle$ and $\mid \downarrow, \rightarrow
\rangle$. Note, that local strain fields result in random effective 
fields in the transverse direction as well. However, their magnitude 
can be estimated to be smaller than that of the TD
interactions\cite{GWT+01}, and more importantly, they are 
symmetrically distributed with zero mean. Their affect on 
the effective transverse field is therefore neglected here.

\begin{figure}
\includegraphics[width = \columnwidth]{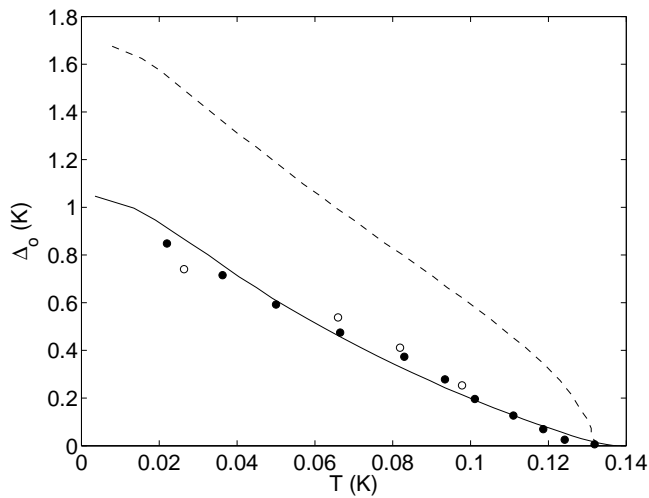}
 \caption{Spin glass - paramagnet phase transition line in the transverse
coupling - temperature ($\Delta_o, T)$ plane, as calculated for the
Hamiltonian in Eq.(\ref{longdipH}) (dashed line) and including the
effective field produced by the off-diagonal dipolar interactions
(solid line). Circles denote experimental data taken from Fig.1 of 
Ref.~[\onlinecite{WBRA93}].  }
 \label{figdiagram}
\end{figure}

In order to demonstrate the significance of the TD 
interaction, we now redo the MF calculation with the simplified
approximation of taking the effective transverse field as
$\tilde{H}_{\perp} = H_{\perp} + H_{\perp}^d$. The effective field due
to the dipolar interaction $H_{\perp}^d$ is the only free parameter in
our calculation. The solid line in Fig.
\ref{figdiagram} is a plot of the SG-PM phase
transition line including the TD interaction, with
$H_{\perp}^d = 0.65$T.\cite{TD} The experimental transition line is
reproduced satisfactorily, including its shape near $T_c$ and the
critical transverse field at low $T$, using a value of $H_{\perp}^d$
that is of the order of the typical effective transverse field
produced by the TD interactions in the system. Note 
that our analysis suggests that it is not sufficient to treat the
TD interactions in MF, i.e., to replace, e.g.,
$J_i^z J_j^x$ by $J_i^z \langle J_j^x \rangle + \langle J_i^z \rangle
J_j^x$.  Only the second term potentially adds to the effective
transverse field, but at the transition there is no such effect in
this approximation since $\langle J_i^z \rangle =0$.

Finally, we consider the reduction of the cusp in the nonlinear
susceptibility $\chi_3(T)$ at low $T$, found experimentally by Wu {\it
et al.} (see Fig. 3 of Ref.~[\onlinecite{WBRA93}]). This result was
interpreted as evidence of a first order phase transition. However,
the experimental results should be reconsidered in view of the
significance of the hf interactions, leading to
Eq.(\ref{IsingHeff}), and of the TD 
interaction. Within the model (\ref{IsingHeff}) the transition is 
expected to be second order.  However, by taking into account the
electro-nuclear nature of the Ising states and the renormalization of
the spin, the diminishing of the cusp is naturally explained:
$\chi_3(T)$ is conjugate to the Edwards-Anderson parameter, and is
proportional \cite{FH91} to $\sum_{ij} \langle S_i S_j \rangle^2 = S_0^4
\sum_{ij} \langle s_i s_j \rangle^2$, where $S_0$ is the effective
electronic spin of the state $\plus$. Experimentally, we assume that
the transition point can be approached with a certain accuracy, giving
a maximum value to $\sum_{ij} \langle s_i s_j \rangle^2$; this would
result in a maximum value $Max (\chi_3(T)) \propto S_0^4$. As $T$ is
reduced $H_c(T)$ increases, thereby reducing $S_0$ and $Max
(\chi_3(T))$, until the cusp can not be experimentally resolved from
the background. In order to establish a different scenario for the
phase transition one must treat properly the TD 
interactions. In Ref.\cite{PTB87} it was shown that a random
longitudinal field smears the SG-PM transition of
the quantum Ising model, and we expect the TD interaction to do the same.

In this letter we have shown that the usual description of the $\LHx$
system, in which both hyperfine and TD interactions
are neglected, is not adequate to explain the SG-PM
phase transition. The large longitudinal hf
interactions of the Ho ion force the relevant Ising doublet to be an
electro-nuclear state, with definite and opposite values of the
spin-$7/2$ nuclear spins. If it were not for transverse hf
interactions, the system would then actually behave like a classical
Ising model with renormalized parameters.  The quantum nature of the
system is revealed only at large transverse fields, proportional to
the energy gap to the first excited electronic state, where quantum
fluctuations induced by the transverse hf interactions become
significant. The TD interaction serves to enhance the
effective transverse field, therefore reducing the critical field of
the transition. A further study of the role of the TD 
interactions would be of interest, since our approximate MF 
treatment does not incorporate multi-spin fluctuations\cite{GTB03}. 
Recently many
interesting phenomena, some connected to the nuclear spins in the
system, were observed in the $\LHx$\cite{BBRA99,JPRA02,RPJ+05}. 
Our approach to the hf
interactions applies to all values of dilution at low $T$, and we hope
it will help in understanding these phenomena as well.

It is a pleasure to thank Gabriel Aeppli, Michel Gingras, Steve
Girvin, Nicolas Laflorencie, Andrea Morello, Subir Sachdev, and Igor
Tupitsyn for useful discussions. This work was supported by NSERC in
Canada, and by PITP.


\begin{thebibliography}{10}
\bibitem{TD}
The transverse dipolar interactions reduce $T_c$ at $H_{\perp}=0$ and we 
therefore set $V_o=0.14$ for $H_{\perp}^d=0.65T$.
\end{thebibliography}
\end{document}